\documentclass[conference]{IEEEtran}
\IEEEoverridecommandlockouts
\usepackage{physics}
\usepackage{balance}
\newcommand{\tsp}
{\mathsf{\scriptstyle{T}}}
\newcommand{\bs}[1]{\boldsymbol{#1}}
\newcommand{\diff}{\ \mathrm{d}}
\newcommand{\R}{\mathbb{R}}

\usepackage{amsthm} % theroemstyle 需要使用的套件

\theoremstyle{definition}

\theoremstyle{definition}

\theoremstyle{definition}

\usepackage{array}
\usepackage{makecell}

\usepackage{blindtext}
\usepackage{xcolor}

\usepackage{cite}
\usepackage{amsmath,amssymb,amsfonts}
\usepackage{algorithmic}
\usepackage{graphicx}
\usepackage{textcomp}
\usepackage{xcolor,soul}
\def\BibTeX{{\rm B\kern-.05em{\sc i\kern-.025em b}\kern-.08em
    T\kern-.1667em\lower.7ex\hbox{E}\kern-.125emX}}

\usepackage{mathrsfs}

\begin{document}
%%---------------------------------
\title{

%Capacity Bounds of Vertically-Drifted First Arrival Position Channels Under Mean-Square Constraint

% Capacity Bounds for Vertically-Drifted First Arrival Position Channels with Second-Moment Constraints

%Capacity Bounds for Vertically-Drifted First Arrival Position Channels under a Second-Moment Constraint

Capacity Bounds for Vertically-Drifted First Arrival Position Channels under a Covariance Constraint

%\iffalse
%{\footnotesize \textsuperscript{*}Note: Sub-titles are not captured in Xplore and
%should not be used}
%\thanks{Identify applicable funding agency here. If none, delete this.}
%\fi
}
%%----
\iffalse
\author{%
  \IEEEauthorblockN{Anonymous Authors}
  \IEEEauthorblockA{%
    Please do NOT provide authors' names and affiliations\\
    in the paper submitted for review, but keep this placeholder.\\
    ISIT23 follows a \textbf{double-blind reviewing policy}.}
}
\fi
%%---------------------------------
\author{
%%---------------------------------
%\and
\IEEEauthorblockN{Yun-Feng Lo}
\IEEEauthorblockA{\textit{Georgia Institute of Technology
} \\
% \textit{School of Electrical and Computer Engineering}\\
Atlanta, Georgia \\
yun-feng.lo@gatech.edu}
%%---------------------------------
\and
\IEEEauthorblockN{Yen-Chi Lee}
\IEEEauthorblockA{\textit{Hon Hai (Foxconn) Research Institute} \\
%\textit{name of organization (of Aff.)}\\
Taipei, Taiwan \\
%yenchilee1925@gmail.com}
yen-chi.lee@foxconn.com}
%%------------------------------
%%------------------------------
\and
\IEEEauthorblockN{Min-Hsiu Hsieh}
\IEEEauthorblockA{\textit{Hon Hai (Foxconn) Research Institute} \\
%\textit{name of organization (of Aff.)}\\
Taipei, Taiwan \\
min-hsiu.hsieh@foxconn.com}
}
\maketitle
%%---------------------------------
\begin{abstract}
\color{black}
In this paper, we delve into the capacity problem of additive vertically-drifted first arrival position noise channel, which models a communication system where the position of molecules is harnessed to convey information. Drawing inspiration from the principles governing vector Gaussian interference channels, we examine this capacity problem within the context of a covariance constraint on input distributions. We offer analytical upper and lower bounds on this capacity for a three-dimensional spatial setting. This is achieved through a meticulous analysis of the characteristic function coupled with an investigation into the stability properties. The results of this study contribute to the ongoing effort to understand the fundamental limits of molecular communication systems.

%Our findings contribute to the broader effort to decipher the fundamental limits of molecular communication systems. They pave the way for further exploration and potential advancements in this cutting-edge field.
\color{black}

%This paper explores the capacity of additive vertically-drifted first arrival position noise channels, in which the position of molecules is used to carry information.
%Borrowing the wisdom from {parallel Gaussian channels}, we investigate this capacity problem subject to a second-moment constraint on input distributions.
%Upper and lower bounds on the capacity are derived for spatial dimension three, based on an analysis of the characteristic function and an investigation of stability properties. The results of this study contribute to the ongoing effort to understand the fundamental limits of molecular communication systems.

\end{abstract}
%%---------------------------------
%%--------------------------------
\section{Introduction}
Molecular communication (MC) is a communication paradigm that employs message molecules (MM) as information carriers \cite{nakano2013molecular,yeh2012new}. Due to its biocompatibility and feasibility at the nanoscale, MC is considered a promising approach for nano-networks \cite{akyildiz2008nanonetworks, farsad2016comprehensive}. In MC systems, MMs act as the information carriers, and a propagation mechanism is necessary to transport them to the receiver. This mechanism can be diffusion-based \cite{jamali2019channel}, flow-based \cite{kadloor2012molecular}, or an engineered transport system such as molecular motors \cite{gregori2010new}. Among these mechanisms, \emph{diffusion-based MC}, often in combination with advection and chemical reaction networks (CRNs), has been the prevalent approach in the literature \cite[Table 4]{farsad2016comprehensive}. 
%We will focus on the study of first-arrival channels in diffusion-based MC systems, including the first arrival time (FAT) and first arrival position (FAP).

%The main advantages of diffusion-based MC include the lack of need for special infrastructure, unlike gap junction-based MC, and no requirement for external energy for the propagation of signaling molecules, unlike motor-based MC. Additionally, the simplicity of diffusion makes it an attractive propagation scheme, particularly for ad hoc networks where mobile nanorobots with limited computational resources form a communication network among themselves and/or with living cells in their close proximity. Therefore, in this paper, we focus on diffusion-based MC, including environments with advection and CRNs.

%The reception mechanism of a receiver can be categorized into two classes: passive reception and active reception \cite{jamali2019channel}. 

\color{black}
The channel characteristics of a diffusive MC system are dependent not only on the physical properties of the propagation medium but also on the reception mechanism. For our study, we adopt the fully-absorbing receiver, which is the common type of active reception \cite{jamali2019channel}. 
%We assume that the receiver can measure both the time and position at which the MM first reaches it, following the assumptions made in \cite{yilmaz2014three,srinivas2012molecular, pandey2018molecular}. 
First arrival models for absorbing receivers can be roughly categorized into three types: time modulation, position modulation, and joint position-time modulation. In the following, we briefly review the progress made in channel capacity research for first arrival time (FAT) and first arrival position (FAP) channels.
\color{black}

\color{black}
FAT channels can be represented as time-invariant additive channels for which
$
T_{\text{out}} = T_{\text{in}} + T_{\text{n}},$
%\label{eq:add-time-ch}
where the arrival time $T_{\text{out}}$ is given by the releasing time $T_{\text{in}}$ plus a random time delay $T_{\text{n}}$ caused by the propagation mechanisms \cite{srinivas2012molecular}. The additive inverse Gaussian noise (AIGN) channel is FAT channel with an inverse Gaussian distribution for the random delay $T_{\text{n}}$. The capacity of the AIGN channel has been studied in the literature, with both upper and lower bounds on the capacity developed \cite{srinivas2012molecular, eckford2012peak, chang2012bounds}. Moreover, \cite{li2014capacity} explores the capacity-achieving input of the AIGN channel, considering both average and peak release time constraints.
\color{black}

\iffalse
Mathematically, first arrival time (FAT) channels can be described as time-invariant additive channels \cite{srinivas2012molecular}:
\begin{equation}
t_{\text{out}} = t_{\text{in}} + t_{\text{n}},
\label{eq:add-time-ch}
\end{equation}
where $t_{\text{in}}$ is the releasing time, $t_{\text{out}}$ is the arrival time, and $t_{\text{n}}$ is the random time delay due to the propagation mechanisms. The authors of \cite{srinivas2012molecular} showed that $t_{\text{n}}$ follows the inverse Gaussian distribution, so this channel is named as the additive inverse-Gaussian noise (AIGN) channel in MC.
Upper and lower bounds on the capacity of the AIGN channel have
been developed in \cite{srinivas2012molecular}. Later, a closeform
upper bound on the capacity of peak constrained AIGN
channel is provided in \cite{eckford2012peak}. Another lower bound on the capacity
of AIGN channel is given in \cite{chang2012bounds}.
The capacity-achieving input
of the AIGN channel with both average and peak release
time constraints is studied in \cite{li2014capacity}.
\fi

%Later, in \cite{srinivas2012molecular, li2014capacity}, capacity bounds for the AIGN channel were derived, and the capacity-achieving input time distribution was also characterized. 
%\color{red}
%(add some more reference about FAT capacity)
%\color{black}

For FAP channels in $\text{D}$-dimensional (dim) spaces, the one-shot channel model can also be written in additive vector form \cite{pandey2018molecular,lee2022arrival} as
\begin{equation}
\bs{X}_{\text{out}} = \bs{X}_{\text{in}} + \bs{X}_{\text{n}},
\label{eq:add-pos-ch}
\end{equation}
where $\bs{X}_{\text{in}}$ is the emission position, $\bs{X}_{\text{out}}$ is the arrival position, and $\bs{X}_{\text{n}}$ is the random position bias incurred by the propagation mechanisms. Note that $\bs{X}_{\text{in}}$, $\bs{X}_{\text{out}}$, and $\bs{X}_{\text{n}}$ are all Euclidean vectors in $\R^{d}$. Throughout this paper, we define $d:=\text{D}-1$.

Although the density function of $\bs{X}_{\text{n}}$ has been derived in previous works \cite{lee2016distribution, pandey2018molecular, lee2022arrival}, an exact characterization of the capacity of FAP channels remains elusive, except for a special case discussed (via numerical simulation) in \cite[Section IV-B]{pandey2018molecular}, where equally spaced $M$-ary modulation is assumed for a 2D FAP channel, and the transition probabilities are discretized. This paper aims to fill this gap by providing analytic capacity bounds for 3D FAP channels without assuming any specific modulation.

%The capacity are explored under a mean-square constraint on the input distributions.

%% --- 主打段落

%\ycl{Analogous} to the parallel Gaussian channels 
%(which are also additive vector channels of the form \eqref{eq:add-pos-ch}) 
%in \cite{cover1999elements}, we

We study the capacity of additive Vertically-Drifted First Arrival Position (VDFAP) noise channels
\begin{align}
\begin{split}
    C 
    = 
    \sup_{f(\bs{X}_\text{in}):~\mathbb{E}\left[ \bs{X}_\text{in}^{\text{ }} \bs{X}_\text{in}^\tsp \right] \preceq \Sigma}
    I(\bs{X}_\text{in};\bs{X}_\text{out})
    ,
\label{eq:def-cap-VDFAP}
\end{split}
\end{align}
where the objective function is the mutual information $I$ between $\bs{X}_\text{in}$ and $\bs{X}_\text{out}$, and the supremum is taken over all input distributions $f(\bs{X}_\text{in})$ satisfying a \emph{covariance matrix constraint} \cite{shang2010capacity} expressed as $\mathbb{E}\left[ \bs{X}_\text{in}^{\text{ }} \bs{X}_\text{in}^\tsp \right] \preceq \Sigma$, where $\Sigma \succ 0$.\footnote{We define the notation $A \succeq B$ as $A-B$ being positive semi-definite, and $A \succ B$ as $A-B$ being positive definite. Similar definitions apply for $\preceq$ and $\prec$.}
To obtain analytic bounds for this capacity, we first derive the characteristic function (CF) of the VDFAP noise distribution. Using the CF, we not only derive formulas for the first two moments but also prove a weak stability property of VDFAP distributions. We then apply the moments and weak stability (see Section~\ref{subsec:stable}) to derive lower and upper bounds for the capacity. 

%This paper is organized as follows:

The structure of this paper is as follows. In Section~\ref{sec:sys}, we present the system model used in our analysis. In Section~\ref{sec:cf}, we examine the characteristic function of the VDFAP distribution. We then use the derived moments and weak stability property to provide lower and upper bounds for the capacity of VDFAP channels in Section~\ref{sec:bounds}. Finally, we summarize our results and provide concluding remarks in Section~\ref{sec:conc}.

%This paper is organized as follows. In Section II, we provide a detailed specification of our system model. The characteristic function of the VDFAP distribution is discussed in Section III. Analytical capacity bounds for VDFAP channels are derived in Section IV. Finally, we conclude our findings in Section V.

%%--------------------------------
\section{System Model}
\label{sec:sys}
%%--------------------------
\newlength{\PLOTWIDTHt}
%%--------------------------
\begin{figure}[!t]
\centering
\includegraphics[width=0.4\textwidth]{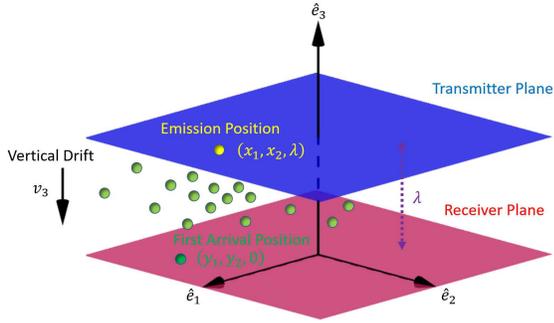}
\caption{
This figure depicts a 3D MC system where both the transmitter and receiver have plane shapes. The communication takes place in a fluid medium with a vertical drift $v_3<0$, and the transmission distance is denoted by $\lambda>0$. The orthonormal basis vectors in 3D space are denoted by $\{\mathbf{\hat{e}}_i\}_{i=1}^3$.
}
\label{fig:dif-cha}
\end{figure}

\color{black}
In this paper, we investigate a diffusion-based molecular communication system, which employs the emission positions of MMs in a $\text{D}$-dimensional fluid medium to convey information. In practice, $\text{D}=2,3$ are commonly used. 
We consider a scenario where there is a constant drift \cite{kadloor2012molecular} in the ambient space. The diffusion effects of MMs are captured by the diffusion coefficient $D=\sigma^2/2$ \cite{farsad2016comprehensive,lee2022arrival}.

Our MC system comprises a transmitter (Tx) and a receiver (Rx), which are modeled as parallel $\text{D}$-dimensional hyperplanes (hereafter referred to as \emph{planes}) separated by a distance $\lambda>0$. Without loss of generality, we can set the transmitter plane at coordinate $x_{\text{D}}=\lambda$ and the receiver plane at coordinate $x_{\text{D}}=0$. Fig.~\ref{fig:dif-cha} demonstrates the 3D case with a vertical drift.

In this study, we assume an ideal MC system that satisfies the following assumptions:
\begin{itemize}
\item The transmitter has perfect control over the emission position of the molecules.
\item The transmitter plane is transparent, allowing MMs to move through it without experiencing any force after they are released.
\item The receiver can perfectly measure the first arrival positions of the molecules.
\item Upon first arrival at the receiver plane, molecules are captured and removed from the system.
\item The movement of every molecule is independent.
\end{itemize}

The positional information is $d$-dim, where $d=\text{D}-1$ for brevity and consistency throughout the paper. The first arrival position $\bs{Y}$ of a molecule released at position $\bs{X}$ can be expressed as $\bs{Y}=\bs{X}+\bs{N}$, where $\bs{N}$ denotes the deviation of the first arrival position. The probability density function of $\bs{N}$ is known as the \emph{FAP density} \cite{lee2022arrival,pandey2018molecular}.

In order to express the FAP density, we denote the $\text{D}$-dim drift vector $\mathbf{v}=[v_1,\ldots,v_\text{D}]^\tsp$ as $\mathbf{v}=[\boldsymbol{v}_\text{par}^\tsp,v_\text{D}]^\tsp$, where $\boldsymbol{v}_\text{par}:=[v_1,\ldots,v_{d}]^\tsp$ contains the drift components parallel to the transmitter and receiver planes, and $v_{\text{D}}$ is the drift component perpendicular to the transmitter and receiver planes. As a rule of thumb, we use bold fonts to stand for column vectors; slanted vectors are $d$-dim, while non-slanted vectors are $\text{D}$-dim. The operator $(\cdot)^\tsp$ represents transposition.
We also introduce the notation
$
\mathbf{u}
:=
\frac{\mathbf{v}}{\sigma^2}
$
and similarly for $\boldsymbol{u}_\text{par}$ and $u_1,\ldots,u_{\text{D}}$. These $\{u_j\}_{j=1}^\text{D}$ can be interpreted as normalized drift.

In \cite{pandey2018molecular,lee2016distribution}, the 2D and 3D FAP densities were separately derived based on the above listed system assumptions.
By the methodology proposed in \cite{lee2022arrival}, we can unify the two FAP densities (i.e., $\text{D}=2,3$) into a single expression: 
\begin{align} \label{eq:dD-FAP}
\begin{split}
    &f^{(d)}_{\bs{N}}(\boldsymbol{n})
    \\=~&
    2\lambda 
    \left(\frac{\| \mathbf{u} \|}{\sqrt{2\pi}}\right)^{d+1} 
    e^{\boldsymbol{u}_{\text{par}}^\intercal \boldsymbol{n}-u_\text{D}\lambda} 
    \frac{K_{\tfrac{d+1}2}\left(\| \mathbf{u} \| \sqrt{\norm{\bs{n}}^2+\lambda^2}\right)}{\left(\| \mathbf{u} \| \sqrt{\norm{\bs{n}}^2+\lambda^2}\right)^{\tfrac{d+1}2}}
    ,
\end{split}
\end{align}
where $K_{\nu}(\cdot)$ denotes the order-$\nu$ modified Bessel function of the second kind \cite{gradshteyn2014table}, and $\|\cdot\|$ denotes the Euclidean norm.

This paper focuses on a sub-family of FAP distributions with two specific properties:
\begin{enumerate}
    \item The \emph{parallel drift} components are zero, i.e., $\boldsymbol{v}_\text{par}=\boldsymbol{0}$.
    \item The \emph{vertical drift} (VD) component points from the Tx to the Rx, i.e., $v_\text{D}<0$. (Intuitively, this vertical drift helps the transmission of information.) 
    %(This vertical drift may aid in the transmission of information, as it helps molecules reach the receiver more quickly.)
\end{enumerate} 
This sub-family is referred to as Vertically-Drifted First Arrival Position (VDFAP) distributions because only the vertical drift component is non-zero. We can set $\boldsymbol{u}_\text{par}=\boldsymbol{0}$ so that $\|\mathbf{u}\|$ in \eqref{eq:dD-FAP} becomes $|u_\text{D}|=-u_\text{D}$.
Abbreviating $u_\text{D}$ as $u$, the VDFAP densities can be expressed as:
\begin{align} \label{eq:dD-VDFAP}
\begin{split}
f^{(d)}_{\bs{N}}(\boldsymbol{n})
=
2\lambda \left(\frac{\vert u \vert}{\sqrt{2\pi}}\right)^{d+1} e^{\lambda\vert u \vert} \frac{K_{\tfrac{d+1}2}\left(\vert u \vert\sqrt{\norm{\bs{n}}^2+\lambda^2}\right)}{\left(\vert u \vert\sqrt{\norm{\bs{n}}^2+\lambda^2}\right)^{\tfrac{d+1}2}}
.
\end{split}
\end{align}
We can thus view $u$ as a parameter of VDFAP distribution.
As a shorthand, we denote $\bs{X}\sim\text{VDFAP}^{(d)}(u,\lambda)$ if a $d$-dim random vector $\bs{X}$ follows the VDFAP distribution defined by \eqref{eq:dD-VDFAP} with parameters $u<0$ and $\lambda>0$.
%Note that in the case of one-dimensional random vectors, we will use the term ``random variables" instead.
\color{black}
Note that we refer to a random variable $X\in\mathbb{R}$ as a ``one-dimensional random vector" to maintain consistency in terminology.
\color{black}
%%--------------------------------
\section{The Characteristic Function of VDFAP Distribution}
\label{sec:cf}

From probability theory, distributions can be characterized by its \emph{characteristic function} (CF). 
The CF viewpoint facilitates the computation of moments and the analysis of stability properties for VDFAP distributions.

For a random vector $\bs{N}$ following $\text{VDFAP}^{(d)}(u,\lambda)$, we denote its CF as
\begin{align}
    \Phi^{(d)}_{\bs{N}}(\boldsymbol{\omega})
    :=
    \mathbb{E}[
    \exp(i\bs{\omega}^\tsp\bs{N})], \text{\ for\ } \boldsymbol{\omega}\in\mathbb{R}^{d}.
\label{eq:CF-def}
\end{align}
We have derived a novel closed-form expression for \eqref{eq:CF-def} when $d=1,2$ as detailed in Appendix~\ref{sec:appen-CF}. The resulting closed-form expression for the CF \eqref{eq:CF-def} is
% The calculation details are presented in Appendix~\ref{sec:appen-CF}. 
% The CF of $\text{VDFAP}^{(d)}(u,\lambda)$ is given by:
\begin{align}
\Phi^{(d)}_{\bs{N}}(\boldsymbol{\omega};u,\lambda)
=
\exp
\left(
-\lambda
\left(
\sqrt{\|\boldsymbol{\omega}\|^2+|u|^2}
-|u|
\right)
\right)
,
\label{eq:VDFAP-CF-revised}
\end{align}
% where $\boldsymbol{\omega}\in\mathbb{R}^{d-1}$ and $|u|$ is the vertical drift.
where we have included the parameters $u,\lambda$ to emphasize the dependency.
\color{black}
It is important to note that letting $|u|\to 0$ in the CF formula \eqref{eq:VDFAP-CF-revised} recovers the $d$-dim Cauchy CF \cite{lee2014clarification,kotz2004multivariate}:
\begin{align}
\lim_{|u|\to 0}
\Phi^{(d)}_{\bs{N}}(\boldsymbol{\omega};u,\lambda)
=
\exp
\left(
-\lambda
\|\boldsymbol{\omega}\|
\right).
\label{eq:Cauchy-CF-revised}
\end{align}
The parameters $u,\lambda$ will be suppressed hereafter when the context allows for clear understanding.
The observation in \eqref{eq:Cauchy-CF-revised} is consistent with \cite{lee2022capacity}, which indicates that as the drift approaches zero, the density of FAP distribution given by \eqref{eq:dD-FAP} reduces to a Cauchy density.
\color{black}

\color{black}
Via the closed-form CF expression \eqref{eq:VDFAP-CF-revised}, we can obtain the first two moments, as well as a weak stability property of VDFAP distributions.
\color{black}

%%--------------------------------
\subsection{Mean Vector and Covariance Matrix}
\color{black}
Based on \eqref{eq:CF-def}, the mean vector of $\bs{N}\sim\text{VDFAP}^{(d)}(u,\lambda)$ can be obtained as
\begin{align}
    \mathbb{E}[\bs{N}]
    =
    -i
    \nabla 
    \Phi^{(d)}_{\bs{N}}(\boldsymbol{\omega})
    \bigg\vert_{\boldsymbol{\omega}=\boldsymbol{0}}
    ,
    \label{eq:VDFAP-mean}
\end{align}
and the correlation matrix of $\bs{N}$ as
\begin{align}
    \mathbb{E}
    \left[
    \bs{N}\bs{N}^\tsp
    \right]
    =
    -
    \nabla^2 
    \Phi^{(d)}_{\bs{N}}(\boldsymbol{\omega})
    \bigg\vert_{\boldsymbol{\omega}=\boldsymbol{0}}
    .
\label{eq:VDFAP-corr}
\end{align}
Due to the radial symmetry of $\bs{N}$, its mean is zero, and thus the correlation matrix coincides with the covariance matrix.
We first calculate the gradient of the CF:
\begin{align}
    \nabla 
    \Phi^{(d)}_{\bs{N}}(\boldsymbol{\omega})
    =
    \frac{-\lambda \Phi^{(d)}_{\bs{N}}(\boldsymbol{\omega})}{\sqrt{\|\boldsymbol{\omega}\|^2+|u|^2}} \boldsymbol{\omega}
    .
\label{eq:VDFAP-CF-grad}
\end{align}
Then we can proceed from \eqref{eq:VDFAP-CF-grad} to calculate the Hessian of the CF:
\begin{align}
\begin{split}
    \nabla^2 
    \Phi^{(d)}_{\bs{N}}(\boldsymbol{\omega})
    &=
    \frac{-\lambda \Phi^{(d)}_{\bs{N}}(\boldsymbol{\omega})}{\sqrt{\|\boldsymbol{\omega}\|^2+|u|^2}}
    \mathbb{I}_{d}
    \\
    &
    ~~
    +
    \frac{\lambda \Phi^{(d)}_{\bs{N}}(\boldsymbol{\omega})}{(\|\boldsymbol{\omega}\|^2+|u|^2)^{\frac32}}
    \left(
    1
    +
    \lambda
    \sqrt{\|\boldsymbol{\omega}\|^2+|u|^2}
    \right)
    \boldsymbol{\omega} \boldsymbol{\omega}^\tsp
    ,
\label{eq:VDFAP-CF-hess}
\end{split}
\end{align}
where $\mathbb{I}_{d}$ denotes the $d\times d$ identity matrix.
Hence, the covariance matrix of $\bs{N}$ can be expressed as:
\begin{align}
    \mathbb{E}\left[ \bs{N}\bs{N}^\tsp \right]
    =
    \frac{\lambda}{|u|} \mathbb{I}_{d}
    .
\label{eq:VDFAP-corr-result}
\end{align}
%Since $\mathbf{N}$ is zero-mean, we also get the covariance matrix of $\mathbf{N}$ (which is the same as its correlation matrix):
\iffalse
\begin{align}
\boldsymbol{\Sigma}_{\mathbf{N}}
    :=
    \mathbb{E}
    \left[
    \left(
    \mathbf{N}-\mathbb{E}[\mathbf{N}]
    \right)
   \left(
    \mathbf{N}-\mathbb{E}[\mathbf{N}]
    \right)
    ^\tsp
    \right]
    =
    \frac{\lambda}{|u|} \mathbb{I}_{d-1}
    .
\end{align}
\fi
\color{black}
It is interesting to note that the variance of each component of a random vector $\bs{N}\sim\text{VDFAP}^{(d)}(u,\lambda)$ is $\frac{\lambda}{|u|}$. Also, the components of $\bs{N}$ are pairwise uncorrelated, but not independent, as the PDF \eqref{eq:dD-VDFAP} cannot be decomposed as the product of its marginal PDFs. 
\color{black}

\subsection{A Weak Stability Property}
\label{subsec:stable}
% We only need the following property for future derivations:

%Cauchy distributions, along with Gaussian and L{\'e}vy distributions, are known to be \emph{stable}. The family of (multivariate) stable distributions is defined to possess a specific form of CF \cite{fallahgoul2013multivariate}. However, as \eqref{eq:VDFAP-CF-revised} does not satisfy this form, we can conclude that VDFAP distributions are \emph{not} stable. Nevertheless, due to their relationship with the Cauchy distribution, VDFAP distributions still exhibit a few ``weaker'' stability properties.

\color{black}
Cauchy distributions, along with Gaussian and L{\'e}vy distributions, are known to be examples of \emph{stable} distributions. These distributions possess a specific form of characteristic function (CF) \cite{fallahgoul2013multivariate}, which is not satisfied by \eqref{eq:VDFAP-CF-revised}. Therefore, we can conclude that VDFAP distributions are \emph{not} stable. However, despite not being strictly stable, VDFAP distributions still exhibit certain ``weaker" stability properties.
\color{black}

\color{black}
One of the weaker stability properties of VDFAP distributions is that if $\bs{N}_1\sim\text{VDFAP}^{(d)}(u,\lambda_1)$ and $\bs{N}_2 \sim\text{VDFAP}^{(d)}(u,\lambda_2)$ are two independent VDFAP random vectors with the same (normalized) drift $u$, then their sum $\bs{N}_1+\bs{N}_2$ follows the $ \text{VDFAP}^{(d)}(u,\lambda_1+\lambda_2)$ distribution.

To prove this property, note that $\bs{N}_1$ and $\bs{N}_2$ are independent. Therefore, the CF of their sum is the product of their respective CFs, i.e.,
$
\mathbb{E}[
    \exp(i\boldsymbol{\omega}^\tsp(\bs{N}_1+\bs{N}_2))]
    =
    \Phi^{(d)}_{\bs{N}_1}(\boldsymbol{\omega})
   \Phi^{(d)}_{\bs{N}_2}(\boldsymbol{\omega}).
$
Using the CF formula \eqref{eq:VDFAP-CF-revised} for $\bs{N}_1$ and $\bs{N}_2$, we obtain:
\begin{align}
\begin{split}
    &\mathbb{E}[
    \exp(i\boldsymbol{\omega}^\tsp(\bs{N}_1+\bs{N}_2))]
    \\=~&
    \exp
            \left(
            -\lambda_1
            \left(
            \sqrt{\|\boldsymbol{\omega}\|^2+|u|^2}
            -|u|
            \right)
            \right)
    \\&
    \cdot
    \exp
            \left(
            -\lambda_2
            \left(
            \sqrt{\|\boldsymbol{\omega}\|^2+|u|^2}
            -|u|
            \right)
            \right)
    \\=~&
    \exp
            \left(
            -\left(\lambda_1+\lambda_2\right)
            \left(
            \sqrt{\|\boldsymbol{\omega}\|^2+|u|^2}
            -|u|
            \right)
            \right) 
    .
\end{split}
\end{align}
Since the CF uniquely characterizes the distribution, we can compare this result to the CF formula \eqref{eq:VDFAP-CF-revised} again to conclude that $\bs{N}_1+\bs{N}_2 \sim \text{VDFAP}^{(d)}(u,\lambda_1+\lambda_2)$.

This \emph{weak stability property} will be useful in deriving a lower bound for the capacity of VDFAP channels.
\color{black}

%%-------------------------------
\section{Bounds on the Capacity of 3D VDFAP Channel}
\label{sec:bounds}

Borrowing the wisdom from vector Gaussian interference channels 
%\cite{cover1999elements}, 
we investigate the VDFAP channel capacity under
the covariance matrix constraint \cite{shang2010capacity,vu2011miso} 
as follows:
\begin{align}
\begin{split}
    C 
    =     \sup_{f(\bs{X}):~\mathbb{E} \left[\bs{X}\bs{X}^\tsp\right] \preceq \Sigma}
    I(\bs{X};\bs{Y}),
\label{eq:def-cap-3D-VDFAP}
\end{split}
\end{align}
where $\Sigma$ is a prescribed \emph{positive definite} matrix.
Note that the channel output $\bs{Y}$ equals to $\bs{X}+\bs{N}$, where the noise $\bs{N}$ follows $\text{VDFAP}^{(2)}(u,\lambda)$ distribution, and is \emph{independent} of the input $\bs{X}$. The capacity appeared in \eqref{eq:def-cap-3D-VDFAP} depends on the parameter triplet $(u,\lambda,\Sigma)$.

%We also use the notation $C(u,\lambda,\Sigma)$ to emphasize its dependency on the parameter triplet $(u,\lambda,\Sigma)$.

% With a derivation similar to \eqref{eq:mutual-info}, we have
Applying the additive channel structure $\bs{Y}=\bs{X}+\bs{N}$, equation \eqref{eq:def-cap-3D-VDFAP} can be simplified as:
\begin{align}
\begin{split}
    C
    =
     \sup_{f(\bs{X}):~\mathbb{E} \left[\bs{X}\bs{X}^\tsp\right] \preceq \Sigma}
    h(\bs{Y}) 
    -
    h(\bs{N})
    ,
\label{eq:cap-3D-VDFAP}
\end{split}
\end{align}
where $h(\cdot)$ denotes the differential entropy. 
%Thus we see that characterizing the differential entropy of VDFAP distributions can be helpful for obtaining bounds on the capacity $C$. 
We have calculated the differential entropy of a random vector $\bs{N}$ following $\text{VDFAP}^{(2)}(u,\lambda)$ in closed-form:
\begin{align}
\begin{split}
    h(\boldsymbol{N}) 
    &=
    \log \left( 2\pi e^3\right)
    + 2\log(\lambda)
    -
    \log (1+\lambda|u|)
    \\
    &~~
    -
    \lambda |u| e^{\lambda |u|}
    \big(
    e \cdot \text{Ei}(-1-\lambda |u|)
    -
    3 \cdot \text{Ei}(-\lambda |u|)
    \big)
    ,
\label{eq:diff-ent-3D-VDFAP}
\end{split}
\end{align}
where the exponential integral function $\text{Ei}(\cdot)$ is defined as \cite{gradshteyn2014table}:
\begin{align}
    \text{Ei}(x)
    :=
    - \int_{-x}^\infty \frac{e^{-t}}{t} 
    \diff t
    , \text{\ for\ } x < 0
    .
\label{eq:def-exp-int}
\end{align}
The calculation of \eqref{eq:diff-ent-3D-VDFAP} is given in Appendix~\ref{sec:appen-DE}.

\subsection{Lower Bound}
Using the weak stability property of VDFAP distributions proved in \ref{subsec:stable}, we obtain a lower bound on the capacity \eqref{eq:def-cap-3D-VDFAP} of the VDFAP channel:
\begin{align}
% \begin{split}
    C(u,\lambda,\Sigma) 
    &\geq
    2 \log \Big( 1 + \tfrac{|u|}{\lambda}\sigma_\text{min} \Big)
    -
    \log \left( 1 + \tfrac{|u|^2}{1+\lambda |u|} \sigma_\text{min} \right)
    \nonumber\\
    &~~
    +
    \lambda |u| e^{\lambda |u|}
    \big(
    e \cdot \text{Ei}(-1-\lambda |u|)
    -
    3 \cdot \text{Ei}(-\lambda |u|)
    \big)
    \nonumber\\
    &~~
    -
    \Big(
    \lambda+|u|\sigma_\text{min}
    \Big)
    |u| \exp\left\{\left(
    \lambda+|u|\sigma_\text{min}
    \right) |u|\right\}
    \nonumber\\
    &~~
    \times
    \bigg\lbrace
    e \cdot 
    \text{Ei}
    \Big(
    -1-
    \big(
    \lambda+|u|\sigma_\text{min}
    \big) 
    |u|
    \Big)
    \nonumber\\
    &~~~~~~
    -
    3 \cdot 
    \text{Ei}
    \Big(
    -
    \big(
    \lambda+|u|\sigma_\text{min}
    \big) 
    |u|
    \Big)
    \bigg\rbrace
    \label{eq:lower-bound} \\
    &>
    0
    \nonumber
    ,
% \end{split}
\end{align}
where $\sigma_\text{min}$ denotes the minimum eigenvalue of $\Sigma$.
\begin{proof}
    To obtain a lower bound on the capacity, we pick particular input distributions $f(\bs{X})$ satisfying the second-moment constraint $\mathbb{E}\left[\bs{X}\bs{X}^\tsp\right]\preceq \Sigma$, calculate $h(\bs{Y})-h(\bs{N})$, and then optimize the result via taking supremum.

    Take $\bs{X}\sim\text{VDFAP}^{(2)}(u,\lambda')$ for some positive $\lambda'\leq |u|\sigma_\text{min}$. By \eqref{eq:VDFAP-corr-result}, we have that 
    $
        \mathbb{E}\left[\bs{X}\bs{X}^\tsp\right]
        =
        \frac{\lambda'}{|u|} \mathbb{I}_2
        \preceq
        \sigma_\text{min} \mathbb{I}_2
        \preceq
        \Sigma
        ,
    $
    so $\bs{X}$ satisfies the 
    covariance constraint. 
    Also, by weak stability property in \ref{subsec:stable}, $\bs{Y}=\bs{X}+\bs{N}$ follows $\text{VDFAP}^{(2)}(u,\lambda+\lambda')$. Then, a lower bound $C \geq h(\bs{Y}) - h(\bs{N}) 
    $ follows from \eqref{eq:cap-3D-VDFAP}.
    % \begin{align}
    %     C \geq h(\bs{Y}) - h(\bs{N}) 
    %     .
    % \label{eq:lower-bound-2}
    % \end{align}
    %$C \geq h(\bs{Y}) - h(\bs{N}).$

    To analyze this lower bound, we introduce an ancillary function $h_0: \R^+ \to \R$ defined by
    \begin{align}
    \begin{split}
        h_0(s)
        &:=
        2 \log (s) 
        - \log (1+s)
        \\
        &\quad
        - s e^s 
        \left(
        e \cdot \text{Ei}(-1-s)
        - 3 \cdot \text{Ei}(-s)
        \right)
        ,\text{\ for\ } s>0,
    \end{split}
    \label{eq:def-h0}
    \end{align}
    so that the differential entropy \eqref{eq:diff-ent-3D-VDFAP} of a random vector $\bs{Z}$, which follows $\text{VDFAP}^{(2)}(u,\lambda'')$, can be written as
    \begin{align}
        h(\mathbf{Z})
        =
        h_0(|u|\lambda'') + \log (2\pi e^3) - 2 \log (|u|)
        .
    \end{align}
    Therefore, the lower bound %\eqref{eq:lower-bound-2} 
    can be expressed via the function $h_0(\cdot)$ as 
    % \begin{align}
    %     C
    %     \geq
    %     h_0(|u|(\lambda+\lambda'))
    %     -
    %     h_0(|u|\lambda)
    %     ,
    % \end{align}
    $    C
        \geq
        h_0(|u|(\lambda+\lambda'))
        -
        h_0(|u|\lambda)
        ,
    $
    which holds for any $0<\lambda'\leq |u|\sigma_\text{min}$. Hence, we can maximize over $\lambda'$ to get 
    % the best lower bound achievable through this technique:
    that
    \begin{align}
        C
        \geq
        \sup_{0<\lambda'\leq |u|\sigma_\text{min}}
        h_0(|u|(\lambda+\lambda'))
        -
        h_0(|u|\lambda)
        .
    \label{eq:lower-bound-3}
    \end{align}
    
    % Numerical simulation (see Fig.~\ref{fig:diff-ent}) shows that the function $h_0(\cdot)$ is strictly increasing. 
    % \begin{figure}[!t]
    %     \centering
    %     \includegraphics[width=0.35\textwidth]{Figures/h_3d_VDFAP.eps}
    %     \caption{
    %     The function $h_0(s)$ versus $s$, which is strictly increasing.
    %     }
    %     \label{fig:diff-ent}
    % \end{figure}
    \color{black}
    It is shown in Appendix \ref{sec:appen-SI} that $h_0(\cdot)$ is strictly increasing.
    \color{black}
    Therefore, \eqref{eq:lower-bound-3} is equivalent to
    \begin{align}
        C
        \geq
        h_0
        \bigg(
        |u|
        \Big(
        \lambda
        +
        |u|\sigma_\text{min}
        \Big)
        \bigg)
        -
        h_0(|u|\lambda)
        .
    \label{eq:lower-bound-2}
    \end{align}
    Plugging \eqref{eq:def-h0} into \eqref{eq:lower-bound-2} yields \eqref{eq:lower-bound}. 
    The lower bound in \eqref{eq:lower-bound-2}, and hence \eqref{eq:lower-bound}, is strictly positive due to the two facts that $h_0(\cdot)$ is strictly increasing, and that $\sigma_\text{min}$ is strictly positive as $\Sigma\succ 0$.
\end{proof}

\subsection{Upper Bound}
It is well-established in the literature that
the differential entropy is maximized by multivariate Gaussian
under a prescribed covariance matrix
\cite[Chapter 12]{cover1999elements},\cite{ebrahimi2008multivariate}.
This fact can be applied to establish an upper bound on the capacity. Under the constraint in \eqref{eq:def-cap-3D-VDFAP}, the bound can be expressed as:
    \begin{align}
    % \begin{split}
        C(u,\lambda,\Sigma) 
        &
        \leq 
        \frac{1}{2} 
        \log
            \left(
            \det
            \Big(
            \tfrac{1}{\lambda^2}\Sigma 
            + 
            \tfrac{1}{\lambda|u|}
            \mathbb{I}_2
            \Big)
            \right)
        +
        \log (1+\lambda|u|)       
        \nonumber\\
        &\quad
        +
        \lambda |u| e^{\lambda |u|}
        \left(
        e \cdot \text{Ei}(-1-\lambda |u|)
        -
        3 \cdot \text{Ei}(-\lambda |u|)
        \right)
        \nonumber\\
        &\quad
        -2,
    \label{eq:upper-bound}
    % \end{split}
    \end{align}
    where $\det(\cdot)$ denotes the determinant.
    \begin{proof}
        Because $\bs{X}$ and $\bs{N}$ are independent and that $\mathbb{E}[ \bs{N} ]=\boldsymbol{0}$,
        \begin{align}
        % \begin{split}
            &
            \mathbb{E}\left[ 
            \bs{Y} \bs{Y}^\tsp 
            \right]
            \nonumber\\
            =~&
            \mathbb{E}\left[ 
            \left( \bs{X} + \bs{N} \right) \left( \bs{X} + \bs{N} \right)^\tsp 
            \right]
            \nonumber\\
            =~&
            \mathbb{E}\left[ 
            \bs{X} \bs{X}^\tsp
            + \bs{X} \bs{N}^\tsp
            + \bs{N} \bs{X}^\tsp
            + \bs{N} \bs{N}^\tsp
            \right]
            \nonumber\\
            \overset{(a)}{=}~&
            \mathbb{E}\left[ 
            \bs{X} \bs{X}^\tsp 
            \right]
            +
            \mathbb{E}\left[ 
            \bs{X} 
            \right]
            \mathbb{E}\left[ 
            \bs{N} 
            \right]^\tsp
            +
            \mathbb{E}\left[ 
            \bs{N} 
            \right]
            \mathbb{E}\left[ 
            \bs{X} 
            \right]^\tsp
            +
            \mathbb{E}\left[ 
            \bs{N} \bs{N}^\tsp 
            \right]
            \nonumber\\
            \overset{(b)}{=}~&
           \mathbb{E}\left[ 
            \bs{X} \bs{X}^\tsp 
            \right]
            +
            \mathbb{E}\left[ 
            \bs{N} \bs{N}^\tsp 
            \right]
            \nonumber\\
            \overset{(c)}{\preceq}~&
            \Sigma 
            + 
            \tfrac{\lambda}{|u|}
            \mathbb{I}_2
            ,
        \label{eq:output-power-constr}
        % \end{split}
        \end{align}
        where (a) is due to independence, (b) is due to $\mathbb{E}[ \bs{N} ]=\boldsymbol{0}$, and (c) is due to the constraint $\mathbb{E}\left[ 
            \bs{X} \bs{X}^\tsp 
            \right] \preceq \Sigma$
        and \eqref{eq:VDFAP-corr-result}.
            
        Under the 
        covariance constraint 
        \eqref{eq:output-power-constr} on the output $\bs{Y}$, its differential entropy $h(\bs{Y})$ is maximized when $\bs{Y}$ follows the bivariate Gaussian distribution $\mathcal{N}
            \Big( 
            \mathbf{0},
            \Sigma 
            + 
            \tfrac{\lambda}{|u|}
            \mathbb{I}_2 
            \Big)$.  
        Hence, from \eqref{eq:cap-3D-VDFAP} we have
        \begin{align}
            C
            \leq
            h
            \bigg( 
            \mathcal{N}
            \Big( 
            \mathbf{0},
            \Sigma 
            + 
            \tfrac{\lambda}{|u|}
            \mathbb{I}_2
            \Big) 
            \bigg)
            -
            h
            \left( 
            \text{VDFAP}^{(2)}(u,\lambda) 
            \right)
            .
        \label{eq:upper-bound-2}
        \end{align}
        Since the differential entropy of the bivariate Gaussian is
        \begin{align}
            h
            \bigg( 
            \mathcal{N}
            \Big( 
            \mathbf{0},
            \Sigma 
            + 
            \tfrac{\lambda}{|u|}
            \mathbb{I}_2
            \Big) 
            \bigg)
            \bigg)
            &=
            % &
            \frac{1}{2}
            \log
            \Big(
            (2\pi e)^2 
            \det
            \big(
            \Sigma 
            + 
            \tfrac{\lambda}{|u|}
            \mathbb{I}_2
            \big)
            \Big)
            % \nonumber\\
            % =~&
            % \log (2\pi e)
            % +
            % \frac{1}{2} 
            % \log
            % \left(
            % \det(\Sigma)
            % +
            % \frac{\lambda}{|u|} \tr (\Sigma)
            % +
            % \frac{\lambda^2}{|u|^2}
            % \right)
            ,
        \end{align}
        together with \eqref{eq:diff-ent-3D-VDFAP} we can simplify \eqref{eq:upper-bound-2} into \eqref{eq:upper-bound}. 
    \end{proof}

\section{Conclusion}
\label{sec:conc}
\color{black}
%The concept of First Arrival Position (FAP) channels has recently emerged in the field of diffusive molecular communication. In this paper, we have contributed to the literature by addressing the capacity problem of additive vertically-drifted (VD) FAP noise channels under a second-moment constraint. 

The concept of First Arrival Position (FAP) channels has been recently introduced in the field of diffusive molecular communication. This paper has advanced the understanding of this novel concept by addressing the capacity problem of additive vertically-drifted (VD) FAP noise channels, specifically under 
the covariance matrix constraint.

The primary contribution of our research lies in providing explicit expressions for both the upper and lower bounds of this capacity as the spatial dimension equals $3$. To arrive at these expressions, we carried out a meticulous analysis of the characteristic function of VDFAP distributions.
Furthermore, our research has uncovered a novel stability property of VDFAP.

Our findings contribute to the ongoing efforts to comprehend the fundamental limits of molecular communication systems. They open new avenues for future research, including extending the analysis to other spatial dimensions and exploring the implications of the discovered stability property.

\bibliographystyle{IEEEtran}
% Your bibliography goes here
\balance
\bibliography{main}
%%--------------------------------
\appendices
\section{Derivation of the Characteristic Function} \label{sec:appen-CF}
%\section{Combined Full Section}

% In this appendix section A, we aim to show that although the VDFAP distribution is not alpha-stable in general, they still enjoy a weaker type of stability. Towards this goal, we first derive the CF of the VDFAP distribution in this subsection and then formulate the weaker stability in the next subsection.

% Notice that PDFs of the VDFAP distribution can be written in a unified form: for $d\in\{2,3\},$ the $d$-dimensional ($d$-dim) VDFAP distribution has the PDF
% \begin{align} 
% \begin{split}
%     f^{(d)}_{\mathbf{N}}(\mathbf{n})
%     &=
%     2\lambda \left(\frac{\vert v_d \vert}{\sqrt{2\pi}}\right)^{d} e^{\lambda\vert v_d \vert} \frac{K_{\frac{d}{2}}\left(\vert v_d \vert\sqrt{\norm{\mathbf{n}}^2+\lambda^2}\right)}{\left(\vert v_d \vert\sqrt{\norm{\mathbf{n}}^2+\lambda^2}\right)^{\frac{d}{2}}}
%     \\&=:
%     f^{(d)}_{\mathbf{N}}(\norm{\mathbf{n}})
%     \text{ for } \mathbf{n}\in\mathbb{R}^{d-1},
% \end{split}
% \end{align}
% where we abused the notation to encode the useful fact that the PDF is radially symmetric; namely, it only depends on $\norm{\mathbf{n}}$. This is a property only present in the VDFAP but not the general FAP, as any non-vertical drift destroys the even symmetry of $\mathbf{N}$ around the origin. Notice that in \eqref{eq:dD-VDFAP} and hereafter in this subsection we assume $v_d<0$ so that this does not degenerate to the case of zero drift. 

% Inspired by the form of \eqref{eq:dD-VDFAP}, w
We apply the Fourier transform (FT) pair relationship between the PDF and CF of Student's $t$-distribution to obtain an integral representation for \eqref{eq:dD-VDFAP}. %(\ref{eq:dD-VDFAP}). 
Specifically, a random variable $N$ following the Student's $t$-distribution of $\nu$ degrees of freedom ($\nu>0$) has the following PDF \cite{gaunt2021simple}:
\begin{align}
    f^{(\nu)}_N(n)
    &=
    \frac{\Gamma\left(\frac{\nu+1}{2}\right)}{\sqrt{\nu \pi} \Gamma\left(\frac{\nu}{2}\right)}
    \left(1+\tfrac{n^2}{\nu}\right)^{-\tfrac{\nu+1}{2}}
    ,\text{\ for\ } 
    n \in \mathbb{R},
\end{align}
where $\Gamma(\cdot)$ is the gamma function, and the corresponding CF:
\begin{align}
    % \Phi^{(\nu)}_N(\omega)
    % &:=
    \mathbb{E}[e^{i\omega N}]
    =
    \frac{K_{\nu/2}(\sqrt{\nu}|\omega|) \cdot(\sqrt{\nu}|\omega|)^{\tfrac{\nu}{2}}}{\Gamma\left(\frac{\nu}{2}\right) 2^{\frac{\nu-2}{2}}}
    ,\text{\ for\ } 
    \omega \in \mathbb{R}
.\end{align}
This implies that for $\omega\in\mathbb{R}\setminus\{0\}$,
\begin{align} \label{eq:Stu-FT-pair}
    K_{\nu/2}(\sqrt{\nu}|\omega|)
    &=
    \frac{2^{\frac{\nu-2}{2}}\Gamma\left(\frac{\nu+1}{2}\right)}{(\sqrt{\nu}|\omega|)^{\frac{\nu}{2}}\sqrt{\nu \pi}}
    \int_{\mathbb{R}}
    \left(1+\tfrac{n^2}{\nu}\right)^{-\tfrac{\nu+1}{2}}
    e^{i\omega n}
    ~\mathrm{d}n
.\end{align}
Fix an arbitrary $s>0$. Applying the substitutions $\tilde{n}=(s/\sqrt{\nu})n$ and $\tilde{\omega}=(\sqrt{\nu}/s)\omega$ to \eqref{eq:Stu-FT-pair}, we have 
\begin{align} \label{eq:K-Int-Rep}
    \frac{K_{\nu/2}(s|\tilde{\omega}|)}{(s|\tilde{\omega}|)^{\frac{\nu}{2}}}
    &=
    \frac{2^{\frac{\nu-2}{2}}\Gamma\left(\frac{\nu+1}{2}\right)}{\sqrt{\pi}|\tilde{\omega}|^{\nu}}
    \int_{\mathbb{R}}
    \left(s^2+\tilde{n}^2\right)^{-\tfrac{\nu+1}{2}}
    e^{i\tilde{\omega}\tilde{n}}
    ~\mathrm{d}\tilde{n}
\end{align}
for $\tilde{\omega}\in\mathbb{R}\setminus\{0\}$. 
%Setting $\nu=d$, $s=\sqrt{\norm{\mathbf{n}}^2+\lambda^2}\ge \lambda>0$ and $\tilde{\omega}=\vert v_d \vert>0$ in \eqref{eq:K-Int-Rep} and plugging into \eqref{eq:dD-VDFAP} gives
Setting $\nu=d+1$, $s=(\norm{\bs{n}}^2+\lambda^2)^{1/2}$ and $\tilde{\omega}=\vert u \vert$ in \eqref{eq:K-Int-Rep}, and then plugging the result into \eqref{eq:dD-VDFAP} yields
\begin{align}
\begin{split}
    f^{(d)}_{\bs{N}}(\bs{n})
    &=
    \frac{\Gamma\left(\frac{d+2}{2}\right)}{\pi^{\frac{d+2}{2}}} \lambda e^{\lambda\vert u \vert} \\
    &\quad\cdot
    \int_{\mathbb{R}} \left(\norm{\bs{n}}^2+\tilde{n}^2+\lambda^2\right)^{-\tfrac{d+2}{2}}
    e^{i\tilde{\omega}\tilde{n}}
    ~\mathrm{d}\tilde{n}. 
\label{eq:dD-VDFAP-PDF-int-rep}
\end{split}
\end{align}
Thus for the $d$-dim VDFAP distribution, its CF
\begin{align}
    \Phi^{(d)}_{\bs{N}}(\boldsymbol{\omega})
    % &:=
    % \mathbb{E}[e^{i\boldsymbol{\omega}\cdot \mathbf{N}}] 
    % \text{ for }
    % \boldsymbol{\omega} \in \mathbb{R}^{d-1}
    % \\
    &=
    \int_{\mathbb{R}^{d}} f^{(d)}_{\bs{N}}(\bs{n})
    ~e^{i\boldsymbol{\omega}^\intercal \bs{n}} 
    ~\mathrm{d}\bs{n}
\label{eq:dD-VDFAP-CF-int-rep}
\end{align}
can be obtained by plugging \eqref{eq:dD-VDFAP-PDF-int-rep} into \eqref{eq:dD-VDFAP-CF-int-rep}, resulting in
\begin{align} \label{eq:CF-FT}
\begin{split}
    \Phi^{(d)}_{\bs{N}}(\boldsymbol{\omega})
    =&
    \frac{\Gamma\left(\frac{d+2}{2}\right)}{\pi^{\frac{d+2}{2}}} \lambda e^{\lambda\vert u \vert}    
     \\
     &\cdot \int_{\mathbb{R}^{d+1}} 
     \big(
     \norm{\mathbf{n}}^2+\lambda^2
     \big)
     ^{-\frac{d+2}{2}}
    e^{
    i\mathbf{w}^\intercal \mathbf{n}
    }
    ~\mathrm{d}\mathbf{n},
\end{split}    
\end{align}
where
\color{black}
$\mathbf{n}:=\left[\bs{n}^\tsp,\tilde{n}\right]^\tsp$ and $\mathbf{w}:=\left[\boldsymbol{\omega}^\tsp,\tilde{\omega}\right]^\tsp$ are $\mathbb{R}^{d+1}$ vectors.
\color{black}
We recognize that the integral in \eqref{eq:CF-FT} is proportional to the CF of a $(d+1)$-variate Cauchy distribution, with the last frequency variable $\tilde{\omega}$ fixed at $\vert u \vert$. The PDF and CF of a $(d+1)$-variate Cauchy vector $\mathbf{X}$ with location $\bs{\mu}=\bs{0}$ and scale $\bs{\Sigma}=\lambda^2 \mathbb{I}_{d+1}$  can be expressed as: \begin{align}
    f_\mathbf{X}(\mathbf{x})
    =
    \frac
    {\Gamma\left(\frac{d+2}{2}\right)}
    { \pi^{\frac{d+2}{2}} }
    \frac{\lambda}{\left( \|\mathbf{x}\|^2 + \lambda^2 \right)^{\frac{d+2}{2}} }
    ,
\end{align}
and
$
    % \Phi_\mathbf{X}(\mathbf{w})
    \mathbb{E}\left[
    e^{i\mathbf{w}^\intercal\mathbf{X}}
    \right]
    =
    \exp
    (-\lambda\|\mathbf{w}\|),
$
as shown in \cite{lee2014clarification,kotz2004multivariate}.
Hence, using the FT pair relationship between the PDF and CF, we can evaluate \eqref{eq:CF-FT} and get
\begin{align}
\begin{split}
    \Phi^{(d)}_{\mathbf{N}}(\boldsymbol{\omega})
    &=
    e^{\lambda|u|} \exp(-\lambda\|\mathbf{w}\|)
    \\&=
    \exp 
    \bigg(
    -\lambda 
    \Big( 
    \sqrt{ \norm{\boldsymbol{\omega}}^2 + \vert u\vert^2 } - \vert u \vert 
    \Big)
    \bigg)
    ,
\end{split}
\end{align}
%where in the last equality we applied $\tilde{\omega}=|u|$.
where we made use of the substitution $\tilde{\omega}=|u|$.

% Based on \eqref{eq:dD-VDFAP-CF}, we can define a special class of $(d-1)$-dim distribution for general dimensions $d\ge 2$. We say that a $(d-1)$-dim random vector $\mathbf{N}$ follows the $\text{VDFAP}(v,\lambda)$ distribution, where $v\ge 0$ (called ``drift") and $\lambda>0$ (called ``distance") are the parameters if its CF takes the form
% \begin{align} \label{eq:VDFAP-CF}
%     \Phi_{\mathbf{N}}(\boldsymbol{\omega})
%     =
%     \exp \left(-\lambda \left( \sqrt{ \norm{\boldsymbol{\omega}}^2 + v^2 } - v \right)\right)
%     \text{ for }
%      \boldsymbol{\omega}\in\mathbb{R}^{d-1}
% .\end{align}
% We further define the set of $(d-1)$-dim VDFAP random vectors as
% \begin{align}
% \begin{split}
%      \mathcal{F}^{(d)}_{\text{VDFAP}}
%     &:=
%     \big\{
%     \text{random vectors}~\mathbf{N}\in\mathbb{R}^{d-1}
%     \big\vert~
%     \\&\quad\quad\quad
%     \exists v\ge 0,\lambda>0~\text{s.t.}~\mathbf{N} \sim \text{VDFAP}(v,\lambda)
%     \big\},
% \end{split}
% \end{align}
% and the set of $(d-1)$-dim VDFAP random vectors with a fixed drift $v\ge 0$ as
% \begin{align}
% \begin{split}
%      \mathcal{F}^{(d,v)}_{\text{VDFAP}}
%     &:=
%     \big\{
%     \text{random vectors}~\mathbf{N}\in\mathbb{R}^{d-1}
%     \big\vert~
%     \\&\quad\quad\quad
%     \exists \lambda>0~\text{s.t.}~\mathbf{N} \sim \text{VDFAP}(v,\lambda)
%     \big\} .
% \end{split}
% \end{align}

%%--------------------------------
\section{Calculation of the 3D Differential Entropy} \label{sec:appen-DE}
From \eqref{eq:dD-VDFAP}, we can write $f^{(2)}_{\bs{N}}(\bs{n}) = c \cdot f(\norm{\bs{n}})$ where
\begin{align}
\begin{split}
    % u
    % &:=
    % \frac{v_3}{\sigma^2}
    % < 0
    % \\
    c 
    &:= 
    \dfrac{\lambda}{\sqrt{2 \pi^3}}|u|^{3} e^{\lambda |u|} 
    ;
    ~
    f(r)
    :=
    \dfrac{K_{3/2}\left(
    |u|\sqrt{r^2+\lambda^2}
    \right)}{\left(|u|\sqrt{r^2+\lambda^2}\right)^{3/2}}
    % \\
    % r
    % &:=
    % \norm{\textbf{n}}
    % =
    % \sqrt{n_1^2+n_2^2}
    .
\label{eq:3DFAP-alternative}
\end{split}
\end{align}
% \begin{align}
% \begin{split}
% f^{(3)}_{\mathbf{N}}(\textbf{n})
% =&\ \dfrac{\lambda}{\sqrt{2 \pi^3}}|v_3|^{3/2} e^{-v_{3} \lambda}
% \dfrac{K_{\frac{3}{2}}\left(
% |v_3|\sqrt{\norm{\textbf{n}}^2+\lambda^2}
% \right)}{\left(\norm{\textbf{n}}^2+\lambda^2\right)^{3/4}}.
% \label{eq:3DFAP-final}
% \end{split}
% \end{align}
Then, the differential entropy of a random vector $\bs{N}$ following $\text{VDFAP}^{(2)}(u,\lambda)$, by definition, is
\begin{align}
    h(\bs{N})
    :=
    - 
    \int_{\mathbb{R}^2}  
    f^{(2)}_{\bs{N}}(\bs{n}) 
    \log 
    \left( 
    f^{(2)}_{\bs{N}}(\bs{n}) 
    \right)
    \mathrm{d}\bs{n}
    .
\end{align}

Since VDFAP distributions are radially symmetric, converting to polar coordinates yields:
\begin{align}
    h(\bs{N})
    =
    - 
    \int_0^{2\pi}
    \int_0^\infty
    c \cdot f(r) \cdot 
    ( \log c + \log f(r) )
    \cdot r
    ~\mathrm{d}r
    ~\mathrm{d}\theta,
\end{align}
where we have applied \eqref{eq:3DFAP-alternative}. Using the fact that $f^{(2)}_{\bs{N}}(\bs{n})$ is a PDF on $\mathbb{R}^2$ and thus integrate to one, we can further express:
\begin{align}
    h(\bs{N})
    =
    -
    \log c
    -
    2\pi c
    \int_0^\infty
    r f(r) \log f(r)
    ~\mathrm{d}r
    .
\label{eq:3DFAP-diff-ent}
\end{align}
Thus, it remains to evaluate the improper integral. Applying the change of variable
\begin{align}
\begin{split}
    \rho
    :=
    |u| \sqrt{r^2+\lambda^2}
    ;~~
    \mathrm{d}\rho
    =
    |u| \frac{r}{\sqrt{r^2+\lambda^2}} \mathrm{d}r
    =
    |u|^2 \frac{r}{\rho} \mathrm{d}r
\end{split}
\end{align}
to this improper integral results in
\begin{align}
\begin{split}
    &\int_0^\infty
    r f(r) \log f(r)
    ~
    \mathrm{d}r
    =
    \int_{|u|\lambda}^\infty
    \frac{K_{3/2}(\rho)}{|u|^2\rho^{1/2}}
    \log
    \left(
    \frac{K_{3/2}(\rho)}{\rho^{3/2}}
    \right)
    \mathrm{d}\rho
    ,
\label{eq:3DFAP-imp-int}
\end{split}
\end{align}
which can be evaluated using the following integral formula: for $a>0$,
\begin{align}
\begin{split}
    &
    \int_a^\infty 
    \frac{K_{3/2}(\rho)}{\rho^{1/2}}
    \log
    \left(
    \frac{K_{3/2}(\rho)}{\rho^{3/2}}
    \right)
    \mathrm{d}\rho
    \\
    =~
    &
    \sqrt{\tfrac{\pi}{2}}
    \Big(
    e 
    \cdot 
    \text{Ei}(-1-a)
    -
    3
    \cdot
    \text{Ei}(-a)
    -
    e^{-a}
    \\
    &
    -
    (2a)^{-1} e^{-a}
    \left(
    6 \log a - 2 \log (1+a) + 6 + \log \tfrac{2}{\pi}
    \right)
    \Big)
    .
\label{eq:3DFAP-int-form}
\end{split}
\end{align}
% where the exponential integral function $\text{Ei}(\cdot)$ is defined as
% \begin{align}
%     \text{Ei}(x)
%     :=
%     - \int_{-x}^\infty \frac{e^{-t}}{t} \mathrm{d}t
%     ~\text{ for }~
%     x < 0
%     .
% \end{align}

% Indeed, the integral formula can be directly verified by taking the derivative with respect to $a$ on both sides and checking they are equal, using the following representation for $K_{3/2}(\cdot)$:
% \begin{equation}
%     K_{3/2}(x)
%     =
%     \sqrt{\dfrac{\pi}{2}}e^{-x}\left(
%     \dfrac{1+x}{x^{3/2}}
%     \right)
%     \text{  for  } x>0
%     .
%     \label{eq:BesselK-one-half}
% \end{equation}
% This will show that the two sides can only differ by a constant. On the other hand, both sides vanish when $a$ approaches infinity, so the constant must be zero.

To further express \eqref{eq:3DFAP-diff-ent} solely in terms of parameters $u$ and $\lambda$, we calculate:
$
    -\log c
    =
    - \log \lambda
    - 3 \log |u|
    - |u| \lambda
    + \frac{1}{2} \log (2\pi^3)
$
and
$
    2\pi c
    =
    \sqrt{\frac{2}{\pi}} \lambda |u|^{3} e^{|u| \lambda}
    .
$
Therefore, taking $a=|u|\lambda$ in \eqref{eq:3DFAP-int-form}, plugging the result into \eqref{eq:3DFAP-imp-int} and then into \eqref{eq:3DFAP-diff-ent} yields
\begin{align}
    h(\bs{N}) &=
    - \log \lambda
    - 3 \log |u|
    - |u| \lambda
    + \tfrac{1}{2} \log (2\pi^3)
    \nonumber\\
    &\quad -
    \lambda |u| e^{|u| \lambda}
    \left(
    e 
    \cdot 
    \text{Ei}(-1-|u| \lambda)
    -
    3
    \cdot
    \text{Ei}(-|u| \lambda)
    \right)
    \nonumber\\
    &
    \quad
    +
    \lambda |u|
    +
    \tfrac{1}{2}
    \left(
    6 \log (|u| \lambda) - 2 \log (1+|u| \lambda)
    \right)
    \nonumber\\
    &\quad
    +
    \tfrac{1}{2}
    \left(
    6 + \log \tfrac{2}{\pi}
    \right)
    % \\
    % &=
    % \log (2\pi e^3)
    % +
    % 2 \log \lambda
    % -
    % \log (1+|u| \lambda)
    % \\
    % &\quad -
    % \lambda |u| e^{|u| \lambda}
    % \left(
    % e 
    % \cdot 
    % \text{Ei}(-1-|u| \lambda)
    % -
    % 3
    % \cdot
    % \text{Ei}(-|u| \lambda)
    % \right)
    .
\end{align}
After collecting terms and simplifying, we arrived at \eqref{eq:diff-ent-3D-VDFAP}.
% Thus the proof is complete.

% \color{blue}
% Here is some more room room room room room room room room room room room room room room room room room room room room room room room room room room room room room room room room room room room room room room room room room room room room room room room room room room room room room room room room room room room room room room room room room room room room room room room room room room room room room room room room room room room room room room room room room room room room room room room room room room room room room room room room room room room room room room room room room room room room room room room room room room room room room room room room room room room room room room room room room.
% \color{black}
%%--------------------------------
\section{Strictly Increasing Property of $h_0(\cdot)$} 
\label{sec:appen-SI}
% To show that the function $h_0(\cdot)$ is strictly increasing, we aim to show that its derivative $h_0'(\cdot)$ is always positive. 

Observe that
$
h_0(s) = 2 \log(s) - \log(1+s) - g(s)
$,
where we introduce a function $g:\R^+ \to \R$ defined by
\begin{align}
    g(s)
    :=
    s e^{s+1} 
    \text{Ei}(-(s+1))
    - 
    3 
    s e^s 
    \text{Ei}(-s),
    \text{  for  } s > 0
    .
\label{eq:def-g}
\end{align}
Taking the derivative of \eqref{eq:def-g} and using the definition of $\text{Ei}(\cdot)$ given in \eqref{eq:def-exp-int}, we obtain the formula
% We can obtain the derivative $g'(\cdot)$ in terms of $g(\cdot)$ using the definition of $\text{Ei}(\cdot)$ given in equation \eqref{eq:def-exp-int}. Specifically, we have:
\color{black}
\begin{align}
g'(s) := \dv{s} g(s)
=
\frac{s+1}{s} g(s) + \frac{s}{s+1} - 3
.
\end{align}
\color{black}
As a consequence, the derivative of $h_0(\cdot)$ can be expressed as
\color{black}
\begin{align}
    h_0'(s) := \dv{s} h_0(s)
    =
    \frac{s+1}{s} (2-g(s))
    .
\end{align}
\color{black}
To show that $h_0'(s)>0$ for any $s>0$, and hence $h_0(\cdot)$ is strictly increasing, it suffices to show that $g(s)<2$ for any $s>0$. Using the inequalities \cite[p.201]{luke1969special}:
\begin{align}
    \frac{s}{s+1} < -s e^s \text{Ei}(-s) < \frac{s+1}{s+2}
    \quad(\forall s > 0)
    ,
\end{align}
we can deduce that
% an upper bound for the function $g(\cdot)$ as
\begin{align}
\begin{split}
    g(s)
    &<
    3 \cdot \frac{s+1}{s+2} - \frac{s}{s+1} \cdot \frac{s+1}{(s+1)+1}
    =
    \frac{2s+3}{s+2}
    <
    2,
    % \text{  for } s > 0
    % ,
\end{split}
\end{align}
which holds for any $s>0$.
Therefore, we have established the previously mentioned sufficient condition for $h_0(\cdot)$ to be strictly increasing.

%%--------------------------------
\end{document}